\documentclass[aps,pra,twocolumn,10pt,showpacs]{revtex4-1}
\usepackage[T1]{fontenc}
\usepackage{amssymb,amsthm,amsmath,amsfonts}
\usepackage{color,hyperref,graphicx,ulem}

\begin{document}

\title{Witnessing random unitary and projective quantum channels:\\ Complementarity between separable and maximally entangled states}

\author{D. Bruns}\affiliation{Institut f\"ur Physik, Universit\"at Rostock, Albert-Einstein-Str. 23, D-18059 Rostock, Germany} %\email{daniel.bruns@uni-rostock.de}
\author{J. Sperling}\email{jan.sperling@uni-rostock.de}\affiliation{Institut f\"ur Physik, Universit\"at Rostock, Albert-Einstein-Str. 23, D-18059 Rostock, Germany}
\author{S. Scheel}\affiliation{Institut f\"ur Physik, Universit\"at Rostock, Albert-Einstein-Str. 23, D-18059 Rostock, Germany} %\email{stefan.scheel@uni-rostock.de}

\date{\today}

\begin{abstract}
	Modern applications in quantum computation and quantum communication require the precise characterization of quantum states and quantum channels.
	In practice, this means that one has to determine the quantum capacity of a physical system in terms of measurable quantities.
	Witnesses, if properly constructed, succeed in performing this task.
	We derive a method that is capable to compute witnesses for identifying deterministic evolutions and measurement-induced collapse processes.
	At the same time, applying the Choi-Jamio\l{}kowski isomorphism, it uncovers the entanglement characteristics of bipartite quantum states. 
	Remarkably, a statistical mixture of unitary evolutions is mapped onto mixtures of maximally entangled states, and classical separable states originate from genuine quantum-state reduction maps.
	Based on our treatment, we are able to witness these opposing attributes at once and, furthermore, obtain an insight into their different geometric structures.
	The complementarity is further underpinned by formulating a complementary Schmidt decomposition of a state in terms of maximally entangled states and discrete Fourier-transformed Schmidt coefficients.
\end{abstract}
\maketitle

%-----------------------------------------------------------------------------------------------
%-----------------------------------------------------------------------------------------------
%-----------------------------------------------------------------------------------------------
\section{Introduction}
	One of the most prominent implications of the quantumness of nature is the existence of nonlocal correlations between compound systems, referred to as entanglement~\cite{EPR35,S35}.
	These kinds of correlations are incompatible with our classical understanding arising from probability theory. 
	For this reason, quantum entangled states are a main resource for applications in quantum computation and quantum communication~\cite{NC00,HHHH09}.
	
	A pure separable state is a product of the form
	\begin{align}\label{eq:Sstate}
		|\psi_\mathrm{S}\rangle=|e\rangle\otimes|f\rangle=|e,f\rangle.
	\end{align}
	Here we assume both subsystems to have identical dimensionality $d$.
	A mixed separable state is defined by statistical mixtures of those pure ones~\cite{W89}.
	Any state that has no such representation is entangled.
	In order to probe the entanglement of a system, experimentally accessible entanglement witnesses have been proposed and optimized~\cite{HHH96,LKCH00,BDHK05,SV09,BHHA13}.
	Another approach has been formulated in terms of so-called positive but not completely positive maps~\cite{HHH96,P96}.

	Entanglement measures have been studied for characterizing the strength of this quantum correlation; cf. Ref.~\cite{HHHH09} for an overview.
	For pure states, the Schmidt decomposition can be used to describe the amount of entanglement~\cite{NC00}.
	By convex roof construction, the so-called Schmidt number has been defined for mixed states and corresponding witnesses have been formulated and optimized~\cite{TH00,SBL01,Betal02,SV11}.
	The standard notion of a maximally entangled (ME) state reads
	\begin{align}\label{eq:MEstate}
		|\psi_\mathrm{ME}\rangle=\frac{1}{\sqrt{d}}\sum_{n=0}^{d-1} |e_n,f_n\rangle,
	\end{align}
	where $\{|e_n,f_{n'}\rangle\}_{n,n'=0,\ldots,d-1}$ is an orthonormal basis, and we have identical Schmidt coefficients $d^{-1/2}$.
	It is important to mention that the definition of an ME state strongly depends on the applied measure~\cite{EP99,MG04,SV11a}.
	However, to be consistent with the literature~\cite{HHHH09}, we will adopt the notion of ME states exclusively for states of the form~\eqref{eq:MEstate}.
	
	Beside the application of quantum correlated states, a determination of the properties of a quantum channel is indispensable.
	Nontrivial applications are, for instance, the description of the propagation of light in turbulent lossy media, such as in atmospheric quantum communication links~\cite{SV10}.
	Quantum channels are also the theoretical foundation for the dynamics of open quantum systems, i.e., the interaction of a system with an environment.
	They may be used for characterizing the Markovian character of a process~\cite{CGLM14,LWHZHLGKPB14} or for formulating the solution of a statistical differential equation (such as Fokker-Planck, Lindblad, or master equations) in quantum physics for a dissipative time evolution~\cite{K72,L76}.

	Apart from their usage in computation and communication protocols, entangled quantum states can also be employed in ancilla-assisted quantum process tomography~\cite{DL01,Aetal03,MMRD03,MRL08}.
	In contrast to standard process tomography~\cite{CN97}, this approach requires only a single bipartite input state to completely characterize an unknown process acting on a quantum system.
	For such a performance, pure ME states~\eqref{eq:MEstate} with perfect quantum correlations turn out to be best suited~\cite{DL03}.
	The underlying idea is given by the Choi-Jamio\l{}kowski isomorphism~\cite{J72,C75}, which provides a one-to-one correspondence between quantum channels and bipartite states,
	that is, the resulting bipartite state characterizes a quantum process completely.
	Hence, properties of quantum operations can directly be linked to their bipartite state representatives~\cite{AP04}.
	A properly constructed witness operator applied to a state representative can uncover properties that rely on a convex structure of the corresponding channels.
	Experimentally accessible witnesses were proposed in Ref.~\cite{MR13}, for instance, in order to detect entanglement-breaking maps or to study separability characteristics of channels.
	Note that we will use the notions channels and maps as synonyms.
	
	Here we derive a technique which allows one to construct witnesses to uncover random unitary (RU) channels and random projective (RP) maps.
	While the former describe a deterministic evolution together with stochastic effects, the latter are based on the quantum state reduction in quantum measurements.
	Applying the Choi-Jamio\l{}kowski isomorphism, RU and RP maps are transformed into mixtures of ME and separable states, respectively.
	For constructing witnesses for such bipartite states or quantum processes, optimization equations are derived to bound the expectation value of general observables.
	Whenever these bounds are exceeded, an RU or RP description or, equivalently, a convex combination of ME or separable states can be excluded.
	The geometric interpretation of a witness as a tangent hyperplane to a given convex set allows one to infer geometric properties of the set itself.
	A complete and full analysis is conducted for arbitrary pure states in a bipartite system along with the introduction of a complementary Schmidt decomposition in terms of ME states and Fourier-transformed Schmidt coefficients.

	The article is structured as follows.
	In Sec.~\ref{sec:channel}, we give the defining relations for RU and RP maps in the context of the Choi-Jamio\l{}kowski isomorphism.
	Structural and geometric properties of both maps are studied.
	The construction of witnesses is performed in Sec.~\ref{sec:witnesses}.
	Witnesses for ME states are derived and compared to witnesses that bound the set of separable states.
	In Sec.~\ref{sec:pure}, the method is applied to perform a full analytical characterization of pure states regarding separability or being ME, e.g., for predicting upper bounds on imperfections.
	The complementary Schmidt decomposition will be defined.
	We summarize and conclude in Sec.~\ref{sec:summary}.

%-----------------------------------------------------------------------------------------------
%-----------------------------------------------------------------------------------------------
%-----------------------------------------------------------------------------------------------
\section{Maximal quantum channels}\label{sec:channel}
	In order to describe the evolution or propagation of physical systems, a process characterization is required; cf. Ref.~\cite{CGLM14} for a recent review on quantum channels.
	For this reason, a convenient form of a linear quantum process is an input-output relation.
	In this form, the initial state of the system $\hat\rho_{\rm in}$ is transformed into a final quantum state $\hat\rho_{\rm out}$,
	\begin{align}\label{eq:In-Out-Rel}
		\hat\rho_{\rm in}\mapsto\hat\rho_{\rm out}=\mathcal E(\hat\rho_{\rm in}).
	\end{align}

	The linear quantum channel $\mathcal E$ itself can be expanded in Kraus operator form~\cite{K83},
	\begin{align}\label{eq:KrausRep}
		\mathcal E(\hat\rho)=\sum_{j}\hat K_j\hat\rho\hat K_j^\dagger.
	\end{align}
	The channels studied here are linear, completely positive (CP) but not necessarily trace preserving.
	The latter property can be restored by properly normalizing the output state of any channel after its application, $\hat\rho_{\rm out}=\mathcal E(\hat\rho_{\rm in})/{\rm tr}[\mathcal E(\hat\rho_{\rm in})]$.

	Besides the Kraus~\cite{K83} and Holevo (not discussed here) representations~\cite{H98}, another key method for characterizing quantum channels is the Choi-Jamio\l{}kowski isomorphism.
	It states that each channel $\mathcal E$ has a unique representation in terms of a bipartite quantum state $\hat\varrho_{\mathcal E}$.
	This isomorphism $\mathcal J$ reads
	\begin{align}\label{eq:CJiso}
		&\mathcal J: \mathcal E\mapsto \hat \varrho_{\mathcal E}=\mathcal N\mathbb I\otimes\mathcal E(|\Phi\rangle\langle \Phi|),
		\\&\text{with }|\Phi\rangle=\sum_{n=0}^{d-1}|n,n\rangle, 
		\label{eq:StdVec}
	\end{align}
	a given normalization constant $\mathcal N$, and a given computational basis $\{|n\rangle\}_{n=0,\dots,d-1}$.
	In the following sections, we highlight two maximal subclasses of maps that will be considered for our further studies.

%-----------------------------------------------------------------------------------------------
\subsection{Random unitary channel}
	Random unitary (RU) channels are characterized by deterministic unitary evolutions $\hat U_j$ which are realized only with a certain probability $p_j$,
	\begin{align}\label{eq:RUdef}
		\mathcal E_{\rm RU}(\hat\rho)=\sum_{j} p_j \hat U_j\hat\rho\hat U_j^\dagger.
	\end{align}
	Such RU maps can be employed to model dephasing that diminishes quantum coherences.
	This allows one to classify decoherence effects~\cite{HS09}, for example, to perform a complete error correction, which is possible if and only if the sources of imperfections are of the RU type~\cite{GW03}.
	These kinds of processes are a main problem to be overcome for the realization of quantum computation~\cite{Z91,CLSZ95}.
	Moreover, RU maps have also been applied to study entanglement dynamics in the presence of environments or phase noise~\cite{ZHHH01,LBAC12,BFACC12}.
	However, except for qubit maps~\cite{LS93}, the full characterization of RU processes remains an open problem.
 
	The Choi-Jamio\l{}kowski isomorphism~\eqref{eq:CJiso} of such random unitary channels yields a convex combination of maximally entangled (ME) states,
	\begin{align}
		\mathcal J(\mathcal E_{\rm RU})=\sum_j p_j|\psi_{\mathrm{ME}, j}\rangle\langle\psi_{\mathrm{ME}, j}|=\hat\varrho_\mathrm{ME},
	\end{align}
	with $\mathcal N=1/d$ and $|\psi_{\mathrm{ME}, j}\rangle=\hat W_j\otimes\hat V_j |\Phi\rangle/\sqrt{d}$ for any pair of unitary maps $\hat W_j,\hat V_j$ satisfying $\hat U_j=\hat V_j\hat W_j^{T}$; cf. Eqs.~\eqref{eq:CJiso} and~\eqref{eq:RUdef}.
	The latter follows from the general relation
	\begin{align}
		\hat A\otimes\hat B |\Phi\rangle=&\hat A\hat B^{T}{\otimes}\hat 1 |\Phi\rangle
		=\hat 1{\otimes}\hat B\hat A^{T} |\Phi\rangle,
		\label{eq:Transfo}
	\end{align}
	where the transposition is taken in the computational basis of the vector $|\Phi\rangle$ given in Eq.~\eqref{eq:StdVec} (see also Appendix~\ref{app:states}).
	Therefore, a characterization of RU channels can be approached by studying ME states~\cite{AS08}.

%-----------------------------------------------------------------------------------------------
\subsection{Random projective channels}
	The complement of the deterministic evolution of quantum states in terms of unitary maps is the highly probabilistic measurement process.
	A measurement of a certain outcome yields the reduction of the state onto the corresponding eigenspace--eigenvectors for nondegenerate observables.
	In classical theories, such a reduction does not occur.
	Hence, the measurement process is a genuine quantum feature.
	Here, we will characterize the corresponding quantum channels.

	Let us consider the following family of maps.
	A CP map is a random projective (RP) channel if it has a Kraus representation of the form
	\begin{align}\label{eq:RPdef}
		\mathcal E_\mathrm{RP}(\hat\rho)=\sum_{j} p_j \hat P_j\hat\rho\hat P_j^\dagger,
	\end{align}
	where $\hat P_j$ describes a rank-one operator, i.e., $\hat P_j=|\phi_j\rangle\langle\psi_j|$, and $\{p_j\}_j$ defines a probability distribution.
	It is worth pointing out that, for finite-dimensional systems, the finite sum is sufficient in this definition due to Carath\'e{}odory's theorem~\cite{C11}.
	In addition, the RP maps are so-called entanglement-breaking channels~\cite{HSR03}.

	In contrast to a unitary evolution in RU channels, an RP map is formulated in terms of collapses of wave functions together with a possible subsequent evolution.
	More rigorously, for each $|\phi\rangle$ there exists a unitary map $\hat U$ such that $|\phi\rangle=\hat U|\psi\rangle$.
	Hence, each term in the RP channel~\eqref{eq:RPdef} can be described as a collapsed state $\hat\rho$ which is further propagated in time,
	\begin{align}\label{eq:singleRP}
		\hat\rho\mapsto\hat U|\psi\rangle\langle\psi|\hat\rho|\psi\rangle\langle\psi|\hat U^\dagger. 
	\end{align}
	Note that such a map is not trace preserving, as $\langle\psi|\hat\rho|\psi\rangle$ describes the (in general) non-unit probability of the reduction to the state $|\psi\rangle\langle\psi|$ within the measurement process.

	The Choi-Jamio\l{}kowski isomorphism $\mathcal J$ in Eq.~\eqref{eq:CJiso} ($\mathcal N=1$) maps entanglement-breaking channels to separable states~\cite{HSR03}.
	Therefore, the RP maps can be identified with the notion of separable states,
	\begin{align}
		\mathcal J(\mathcal E_\mathrm{RP})
		=&\sum_{j}p_j|\psi_j^\ast\rangle\langle \psi_j^\ast|\otimes|\phi_j\rangle\langle\phi_j|=\hat\varrho_\mathrm{S},
	\end{align}
	where we used the relation
	\begin{align}
		(\hat 1\otimes\langle\psi|)|\Phi\rangle{=}\sum_{m=1}^d |m\rangle\langle\psi|m\rangle{=}\sum_{m=1}^d |m\rangle \langle m|\psi \rangle^\ast{=}|\psi^\ast\rangle,
	\end{align}
	and $|\psi^\ast\rangle$ is the complex conjugate of the vector $|\psi\rangle$ in the computational basis.
	The state $\hat\varrho_\mathrm{S}$ describes a separable state~\cite{W89}, and any pure state, given as $|\psi^\ast\rangle\langle\psi^\ast|\otimes|\phi\rangle\langle\phi|$, can be obtained from the RP map~\eqref{eq:singleRP} with a single element.
	Hence, an identification of an RP channel is equivalent to the separability problem.

%-----------------------------------------------------------------------------------------------
\subsection{Maximal states and CP maps}
\begin{figure}
		\includegraphics[width=7.5cm]{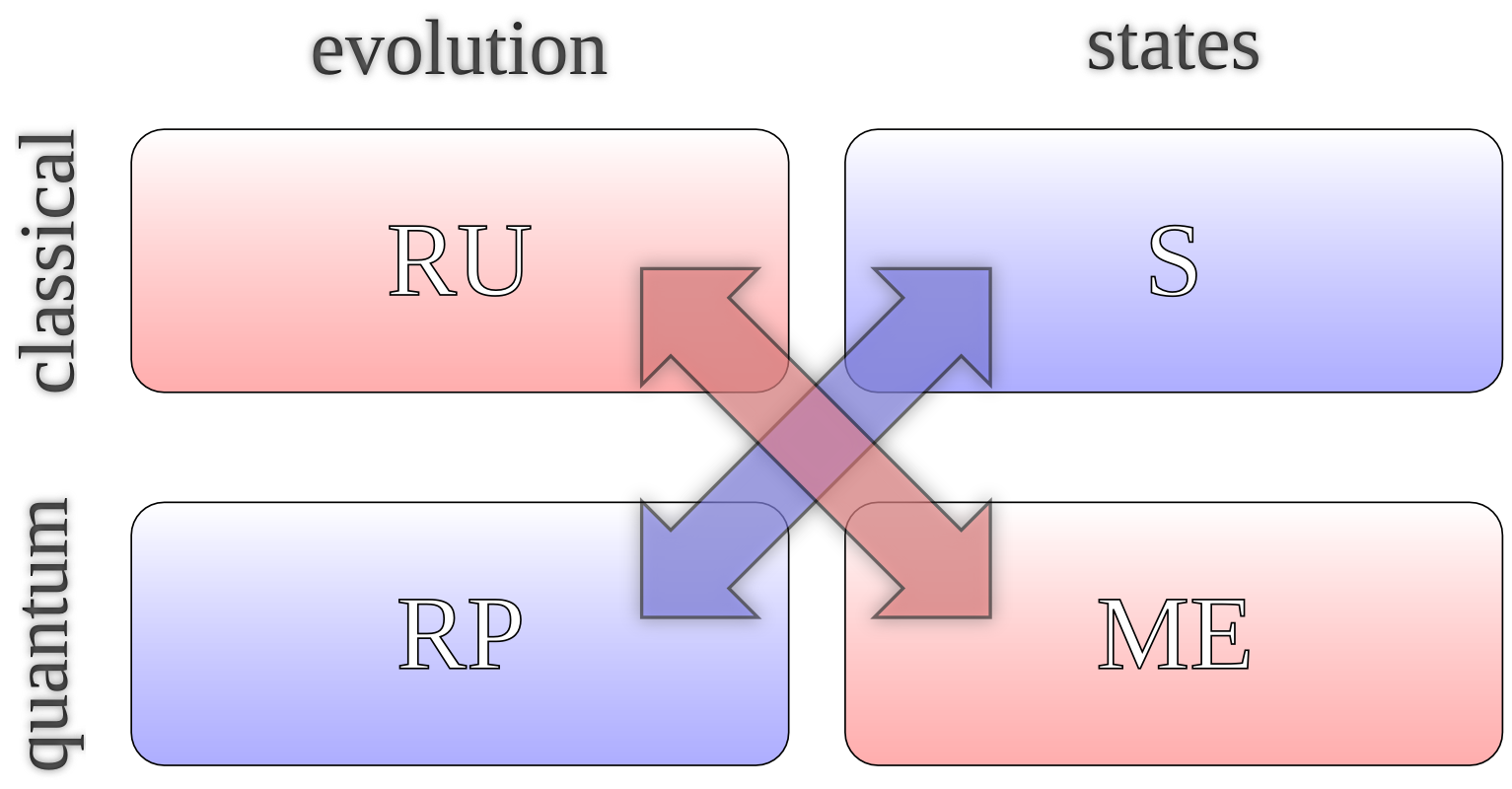}
		\caption{(Color online)
			The mapping of the Choi-Jamio\l{}kowski isomorphism $\mathcal J$ is depicted via the arrows.
			On the one hand, the deterministic evolution together with a classical statistical description (RU maps) is mapped onto the set of bipartite states which are formed by pure entangled states with equally weighted Schmidt coefficients, i.e., ME states.
			On the other hand, the maps which correspond to the genuine quantum description of the measurement process (RP maps) have the image of the set of classically correlated (separable) states.
		}\label{fig:statesVSmaps}
	\end{figure}

	Let us summarize some initial observations.
	Using the isomorphism $\mathcal J$, it was shown that the problem of identifying specific kinds of quantum channels can be mapped onto the characterization of bipartite states.
	A pictorial summary may be found in Fig.~\ref{fig:statesVSmaps}.

	First, we recalled the fact that RU maps have a bipartite representation in terms of mixtures of ME pure states~\cite{AS08}.
	From now on, we will use the notion ME state for such mixed and pure states, even though some of the mixtures are separable (for example, the normalized identity; cf. Appendix~\ref{app:Fourier}).
	We emphasize that such a deterministic evolution is something one also expects for a classical channel.
	The image of $\mathcal J$, however, is the convex hull of pure ME states and those pure ME states have genuine quantum correlations between the subsystems.

	Second, we established the set of RP maps.
	The physical interpretation of such maps is a measurement-induced state reduction.
	Again, we stress that this aspect of quantum physics has no counterpart in the classical domain.
	The action of $\mathcal J$ behaves in a complementary way as it yields bipartite separable states sharing no quantum entanglement. 

	Hence, there is a cross correlation between states and channels (Fig.~\ref{fig:statesVSmaps}).
	Nonclassical RP and classical RU maps are propagated to separable and ME states, respectively.
	Due to the fact that $\mathcal J$ is a bijective transformation, we focus on the determination of separable and ME states from now on.
	However, one should keep in mind for the remainder of this work that one can draw all of the following conclusions for the corresponding channels.

%-----------------------------------------------------------------------------------------------
\subsection{Geometric representation}\label{subsec:geoState}
	Let us consider some geometric aspects of the set of separable states and ME states.
	The respective extremal points are pure separable states~\eqref{eq:Sstate} and pure ME states~\eqref{eq:MEstate}.
	In general, the convex set of all bipartite quantum states is convexly spanned by all pure states, each having a distinct Schmidt decomposition~\cite{NC00}:
	\begin{align}\label{eq:SchmidtDec}
		|\psi\rangle=\sum_{n=0}^{d-1}\sigma_n|e_n,f_n\rangle,
	\end{align}
	where $\sigma_n$ is the $n$th non-negative Schmidt coefficient.

	For the time being, let us restrict ourselves to the family of pure states $\{|\psi^{(j)}\rangle\}_j$ having a decomposition with identical vectors $\{|e_n,f_n\rangle\}_{n=0,\dots,d-1}$ but different Schmidt coefficients $\sigma^{(j)}_n$.
	The convex span of those pure states is $\mathcal C=\mathrm{conv}\{|\psi^{(j)}\rangle\langle\psi^{(j)}|\}_j$.
	For the spanned mixed states, $\hat\rho=\sum_j p_j |\psi^{(j)}\rangle\langle \psi^{(j)}|\in\mathcal C$, we define the following projections:
	\begin{align}
		\sigma_n^2=\langle e_n,f_n|\hat\rho^2|e_n,f_n\rangle=\sum_{m=0}^{d-1} \left(\sum_j p_j \sigma^{(j)}_{m}\sigma^{(j)}_{n}\right)^2.
	\end{align}
	For the considered class of pure states, these definitions of $\sigma_n^2$ coincide with the squares of Schmidt coefficients.
	In general, the purity yields ${\rm tr}(\hat\rho^2)=\sum_{n=0}^{d-1}\sigma_n^2\leq 1$.

	For the subspace $\mathcal C$, it holds that it is the convex hull of states satisfying $\sum_{n=0}^{d-1}\sigma_n^2=1$.
	Similarly to the Bloch-sphere representation, we obtain the full ball of pure and mixed quantum states from this high-dimensional sphere.
	In fact, one finds only one hyperoctant of the sphere.
	Hence, for symmetry reasons, we may allow $\sigma_n^{(j)}<0$ for pure states or $\sigma_n=\pm[\sigma_n^2]^{1/2}$ for mixed ones.
	Using the vector representation $\vec \sigma=(\sigma_0,\dots,\sigma_{d-1})^T\in\mathbb R^d$, we can alternatively describe the ball as $\|\vec \sigma\|_2=[\sum_{n=0}^{d-1}|\sigma_n|^2]^{1/2}\leq 1$.
	A state in the considered subspace given by Eq.~\eqref{eq:SchmidtDec} is pure if and only if $\|\vec \sigma\|_2=1$.

	In this form, a separable pure state is characterized by all points on the sphere ($\|\vec \sigma\|_2=1$) where one and only one Schmidt coefficient is nonvanishing, $|\sigma_{n_0}|=1$, for a given $n_0$.
	This means that a pure state is separable if and only if $\|\vec \sigma\|_2=1$ and $\|\vec\sigma\|_1=1$.
	The enclosed convex volume defines the hyperdimensional analog to an octahedron.
	In vector notion, this set is given by $\|\vec\sigma\|_1\leq 1$, with the $1$-norm $\|\vec \sigma\|_1=\sum_{n=0}^{d-1}|\sigma_n|$.

	For pure ME states, all Schmidt coefficients have the same magnitude, $|\sigma_0|=\dots=|\sigma_{d-1}|=d^{-1/2}$.
	This is equivalent to the intersection of vectors $\vec \sigma$ which satisfy $\|\vec \sigma\|_2=1$ and $\|\vec \sigma\|_\infty=d^{-1/2}$ simultaneously.
	The convex combination of these vertices yields a hypercube, $\|\vec\sigma\|_\infty\leq d^{-1/2}$, by applying the maximum norm $\|\vec\sigma\|_\infty=\max\{|\sigma_n|\}_{n=0,\dots,d-1}$.

\begin{figure}
		\includegraphics[width=7.5cm]{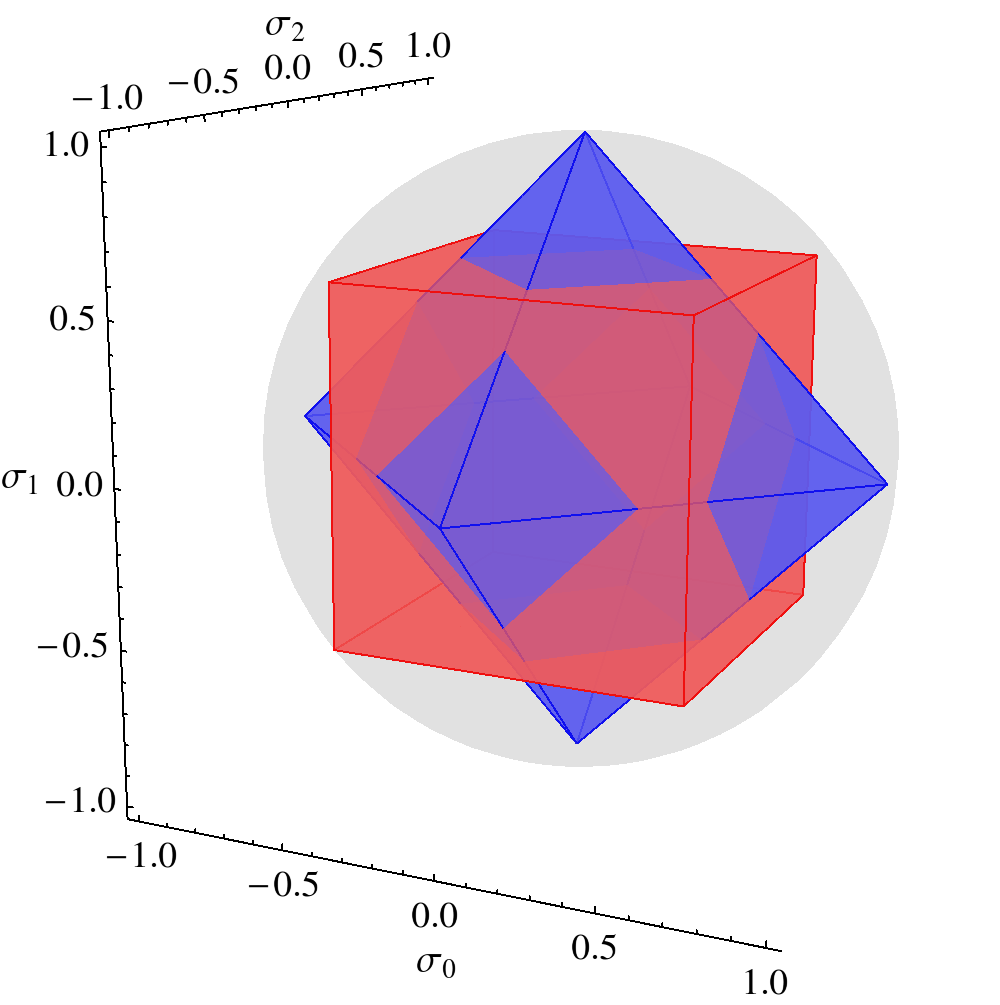}
		\caption{(Color online)
			The (gray) ball depicts the volume of all mixed quantum states which is bounded by states of the form~\eqref{eq:SchmidtDec} for $d=3$.
			The octahedron (blue) represents the set of separable states and the cube (red) describes ME states.
		}\label{fig:convex}
	\end{figure}

	In Fig.~\ref{fig:convex}, the three-dimensional case is shown.
	Note that for finite-dimensional systems the normed spaces defined by $\|\,\cdot\,\|_1$ and $\|\,\cdot\,\|_\infty$ are dual to one another~\cite{Y08}.
	This also highlights the complementary relations between separable and ME states.
	A similar relation between entanglement witnesses and separable states in two-qubit systems was reported recently~\cite{MJR15}.
	Moreover, a numerical study in Ref.~\cite{HM15} was performed for a similar, i.e., geometric, characterization of positive but not completely positive maps.
	Local properties of quantum channels and their verification have been further studied in Ref.~\cite{MR13}.

%-----------------------------------------------------------------------------------------------
%-----------------------------------------------------------------------------------------------
%-----------------------------------------------------------------------------------------------
\section{Witnesses for maximally entangled and separable states}\label{sec:witnesses}
	In this section, we derive observable conditions which enable us to infer whether or not a state is an ME state.
	This will result in nonlinear eigenvalue equations whose solutions give the upper or lower bound of an observable for the desired class of states.
	Eventually, we will compare our method with the construction of entanglement witnesses.

%-----------------------------------------------------------------------------------------------
\subsection{Witness construction}
	In order to formulate witnesses for ME states, let us apply the Hahn-Banach separation theorem~\cite{Y08,M33}.
	It states that for any closed, convex set and any element that is not part of this set, there exists a linear functional that separates the element from the set.
	In our case, the closed convex set is the set of mixtures of ME states.
	Any linear functional $f$, acting on trace class operators $\hat\rho$, can be written as $f(\hat\rho)={\rm tr}(\hat\rho\hat L)$ for a bounded, Hermitian operator $\hat L$.
	The separation of a non-ME state $\hat\varrho$ in a finite-dimensional system reads as follows.
	There exists a Hermitian operator $\hat L$, such that
	\begin{align}\label{eq:MEwitness}
		\langle \hat L\rangle>\max\{f(\hat\rho_{\rm ME})\}_{\hat\rho_{\rm ME}}=g_{\rm ME}^{\max}.
	\end{align}
	The value of the functional, $\langle \hat L\rangle={\rm tr}(\hat\varrho\hat L)$, corresponds to an experimentally accessible expectation value of the observable $\hat L$.
	Due to convexity, the maximal expectation value for ME states, $g_{\rm ME}^{\max}$, is attained for a pure state.
	Thus, this bound can be formulated in terms of an optimization over pure ME states $|\psi_{\rm ME}\rangle$,
	\begin{align}\label{eq:OptProbl}
		g_{\rm ME}=\langle\psi_{\rm ME}|\hat L|\psi_{\rm ME}\rangle\to g_{\rm ME}^{\max}.
	\end{align}

	Recall that any ME pure state can be written as
	\begin{align}
		|\psi_\mathrm{ME}\rangle=\frac{1}{\sqrt d}\hat 1\otimes\hat U|\Phi\rangle=\frac{1}{\sqrt d}\sum_{n=0}^{d-1}|n,u_n\rangle
	\end{align}
	[see Eq.~\eqref{eq:Transfo} or Appendix~\ref{app:states}], together with orthonormality constraints for $\{|u_n\rangle\}_{n=0,\dots,d-1}$ of the form
	\begin{align}\label{eq:Constaints}
		c_{i,j}=\langle u_i|u_j\rangle-\delta_{i,j} \equiv 0,
	\end{align}
	for $i,j=0,\ldots,d-1$ and the Kronecker symbol $\delta_{i,j}$.
	Additionally, let us decompose the observable $\hat L$ into the computational basis of the first subsystem,
	\begin{align}
		\hat L=\sum_{i,j=0}^{d-1} |i\rangle\langle j|\otimes \hat L_{i,j},
	\end{align}
	which yields
	\begin{align}
		g_{\rm ME}=\frac{1}{d}\sum_{i,j=0}^{d-1} \langle u_i|\hat L_{i,j}|u_j\rangle.
	\end{align}
	Now, the optimization problem~\eqref{eq:OptProbl} under the constraints~\eqref{eq:Constaints} can be solved by the method of Lagrange's multipliers $\gamma_{i,j}$.
	That is, for all $k=0,\ldots, d-1$, we have a vanishing gradient of the form
	\begin{align}
		0{=}&\frac{\partial g_{\rm ME}}{\partial \langle u_k|}{-}\sum_{i,j=0}^{d-1}\gamma_{i,j}\frac{\partial c_{i,j}}{\partial \langle u_k|}
		{=}\frac{1}{d}\sum_{j=0}^{d-1} \hat L_{k,j}|u_j\rangle{-}\sum_{j=0}^{d-1}\gamma_{k,j}|u_j\rangle,
	\end{align}
	as $\partial \langle u_i|/\partial \langle u_k|=\delta_{i,k}$.
	It can be checked by a projection onto $\langle k|$ in the first subsystem that we can write equivalently
	\begin{align}
		\hat L|\psi_{\rm ME}\rangle=d(\hat 1\otimes \hat \Gamma)|\psi_{\rm ME}\rangle,
		\label{eq:ME-EValEq}
	\end{align}
	with $\hat\Gamma=\sum_{i,j=0}^{d-1} \gamma_{i,j} |u_j\rangle\langle u_i|$.
	In this form, we have the generalized eigenvalue problem~\eqref{eq:ME-EValEq} with an ME eigenstate $|\psi_{\rm ME}\rangle$.
	The corresponding generalized eigenvalue, denoted as $g_{\rm ME}^{\rm opt}$, is given by
	\begin{align}
		\nonumber g_{\rm ME}^{\rm opt}{=}&\langle\psi_{\rm ME}|\hat L|\psi_{\rm ME}\rangle{=}d\langle\psi_{\rm ME}|\hat 1{\otimes}\hat \Gamma|\psi_{\rm ME}\rangle
		{=}\sum_{i=0}^{d-1} \langle u_i|\hat\Gamma|u_i\rangle
		\\{=}&{\rm tr}(\hat\Gamma).
	\end{align}
	Finally, the maximal expectation value of $\hat L$ for ME states is given as the maximum over all optimal values,
	\begin{align}\label{eq:maxMEvalue}
		g_{\rm ME}^{\max}=\max\{g_{\rm ME}^{\rm opt}\},
	\end{align}
	which is the desired right-hand side of the ME test in inequality~\eqref{eq:MEwitness}.

	The value of $g_{\rm ME}^{\max}$ in Eq.~\eqref{eq:maxMEvalue} is a tight upper bound, as it is attained for the corresponding eigenvector $|\psi_{\rm ME}\rangle$ solving Eq.~\eqref{eq:ME-EValEq}, which exists.
	This is due to the fact that all pure ME states form a bounded and closed subset of the finite-dimensional and, therefore, compact unit sphere of normalized pure states~\cite{Y08}.
	It is worth mentioning that the same procedure can be performed similarly for a minimum.
	That is, $\hat\varrho$ is not an ME state if and only if there exists an observable $\hat L$ such that
	\begin{align}
		\langle\hat L\rangle<g_{\rm ME}^{\min}=\min\{g_{\rm ME}^{\rm opt}\},
	\end{align}
	which can be deduced from the approach with the maximum via the interchange $\hat L\mapsto-\hat L$.

%-----------------------------------------------------------------------------------------------
\subsection{Relation to other eigenvalue problems}
	Computing the upper bound for all quantum states can be done by solving the (standard) eigenvalue problem for finding the maximal eigenvalue.
	This is consistent with our finding that the generalized eigenvalue problem in Eq.~\eqref{eq:ME-EValEq} yields the upper bound for ME states.
	At this point, we can determine what might be a useful witness.
	For example, a witness based on $\hat L$ for which $g_{\rm ME}^{\max}$ is the ultimate upper bound to all quantum states cannot fulfill condition~\eqref{eq:MEwitness}.
	In general, we can make the following statement.
	The observable $\hat L$ is a proper witness if and only if the eigenspace to the (standard) maximal eigenvalue does not contain an ME state.
	The proof is straightforward:
	
	First, as we pointed out above, if the eigenspace contains an ME state $|\psi_{\rm ME}\rangle$, $g_{\rm ME}^{\max}=\langle\psi_{\rm ME}|\hat L|\psi_{\rm ME}\rangle$ is identical to the maximal eigenvalue.
	Thus, condition~\eqref{eq:MEwitness} is empty.
	Secondly, if the eigenspace does not contain such an ME state, any element $|\psi\rangle$ of the eigenspace to the maximal (standard) eigenvalue will satisfy the test~\eqref{eq:MEwitness}, $\langle \psi|\hat L|\psi\rangle>g_{\rm ME}^{\max}$.
	A similar statement has been formulated for entanglement and so-called Schmidt number witnesses~\cite{SV11}.

	Moreover, in order to prove that $\hat\varrho$ is not a separable state, a similar relation to~\eqref{eq:MEwitness} has been derived~\cite{SV09},
	\begin{align}\label{eq:entWitness}
		\langle\hat L\rangle>g_{\rm S}^{\max},
	\end{align}
	with $g_{\rm S}^{\max}=\max\{g_{\rm S}^{\rm opt}\}$.
	The latter values are determined from the so-called separability eigenvalue equations,
	\begin{align}\label{eq:SEValEq}
		\hat L_b|a\rangle=g_{\rm S}^{\rm opt}|a\rangle
		\text{ and }
		\hat L_a|b\rangle=g_{\rm S}^{\rm opt}|b\rangle,
	\end{align}
	with $\hat L_a={\rm tr}_A[\hat L(|a\rangle\langle a|\otimes \hat 1)]$, $\hat L_b={\rm tr}_B[\hat L(\hat 1\otimes|b\rangle\langle b|)]$, and $\langle a|a\rangle=1=\langle b|b\rangle$.
	This kind of approach has been used to experimentally uncover path-entangled states~\cite{Getal14} or for studying entanglement from semiconductor systems~\cite{PFSV12}.

	For separable states, the generalized eigenvalue problem in Eq.~\eqref{eq:SEValEq} has the same meaning as Eq.~\eqref{eq:ME-EValEq} for ME states.
	However, the corresponding maximal bounds $g_{\rm S}^{\max}$ and $g_{\rm ME}^{\max}$ address the detection of different properties.
	On the one hand, if condition~\eqref{eq:entWitness} is fulfilled, then the state $\hat\varrho$ is entangled.
	If, on the other hand, Eq.~\eqref{eq:MEwitness} is fulfilled, then $\hat\varrho$ is not an ME state and it is thus not the Choi-Jamio\l{}kowski state of an RU process.

%-----------------------------------------------------------------------------------------------
\subsection{On computing solutions}
	In the following, let us solve Eq.~\eqref{eq:ME-EValEq} for some classes of operators in order to demonstrate the functionality of our method.
	The solutions will yield measurable tests in Eq.~\eqref{eq:MEwitness} to probe ME states.
	Our results will be compared with known tests for RU maps and related to those for entanglement detection.

\subsubsection{Product operators}
	As a first example, we consider a simple correlation measurement between the two modes.
	Let
	\begin{align}
		\hat L=\hat A\otimes \hat B
	\end{align}
	be a Hermitian positive semidefinite operator.
	Inserted into Eq.~\eqref{eq:ME-EValEq}, we find
	\begin{align}
		\hat L[\hat 1\otimes\hat U]|\Phi\rangle=d[\hat 1\otimes\hat \Gamma][\hat 1\otimes\hat U]|\Phi\rangle,
	\end{align}
	with $|\psi_\mathrm{ME}\rangle=\hat 1\otimes\hat U|\Phi\rangle/\sqrt{d}$.
	This gives
	\begin{align*}
		[\hat A\otimes\hat B\hat U]|\Phi\rangle=&[\hat 1\otimes(\hat B\hat U\hat A^T)]|\Phi\rangle=[\hat 1\otimes (d\hat\Gamma\hat U)]|\Phi\rangle.
	\end{align*}
	Equating coefficients yields
	\begin{align}
		\hat\Gamma=\frac{1}{d}\hat B\hat U\hat A^T\hat U^\dagger
		\text{ and }
		g^{\rm opt}_{\rm ME}=\frac{1}{d}{\rm tr}(\hat B\hat U\hat A^T\hat U^\dagger).
	\end{align}
	The spectral decomposition of the considered product observable reads as $\hat L=\sum_{m=0}^{d-1}\lambda_{A,m} |a_m\rangle\langle a_m|\otimes\sum_{n=0}^{d-1}\lambda_{B,n} |b_n\rangle\langle b_n|$, with eigenvalues sorted in increasing order.
	Using this fact, its positive semidefiniteness, and Chebyshev's sum inequality (see Appendix~\ref{app:Cheb}), we have
	\begin{align}
		g_{\rm ME}^{\max}{=}\frac{1}{d}\sum_{n=0}^{d-1}\lambda_{A,n}\lambda_{B,n}
		\text{ and }
		g_{\rm ME}^{\min}{=}\frac{1}{d}\sum_{n=0}^{d-1}\lambda_{A,n}\lambda_{B,d{-}1{-}n}.
	\end{align}

	In the case of separable states, we can deduce from the solution of the separability eigenvalue problem~\eqref{eq:SEValEq} that
	\begin{align}
		g_{\rm S}^{\max}=\lambda_{A,d-1}\lambda_{B,d-1}
		\text{ and }
		g_{\rm S}^{\min}=\lambda_{A,0}\lambda_{B,0}.
	\end{align}
	Comparing these values with the spectral decomposition of $\hat L$, we find that such a product operator cannot be a proper entanglement witness because the upper and lower bounds for all states are identical with those for separable states.
	However, for nontrivial scenarios, they differ from the $g_{\rm ME}^{\max{/}\min}$.
	Hence, such a correlation measurement is a proper witness to identify states which cannot be a mixture of ME states, or non-RU maps.

	An interesting consequence of such witnesses is given by $\hat A=\hat 1$.
	In this case, the upper and the lower bound coincide: $g_{\rm ME}^{\max}={\rm tr}(\hat B)/d=g_{\rm ME}^{\min}$.
	In terms of expectation values, this means that a violation of $\langle \hat L\rangle={\rm tr}_B[\hat B{\rm tr}_A(\hat\rho)]={\rm tr}(\hat B)/d$ for arbitrary $\hat B$ identifies a non-ME state.
	This simple consequence of our technique is equivalent to a previously known constraint onto mixtures of ME states~\cite{AS08,AP04}:
	\begin{align}\label{eq:unital}
		{\rm tr}_A(\hat\rho_{\rm ME})=\hat 1/d.
	\end{align}
	In terms of the isomorphism $\mathcal J$ in Eq.~\eqref{eq:CJiso}, this means that the violation of Eq.~\eqref{eq:unital} excludes the RU description of the channel.
	The constraint is clearly violated, for instance, for pure separable states or the projective channel in Eq.~\eqref{eq:singleRP}.
	A similar treatment for $\hat B=\hat 1$ gives the same restriction for the other subsystem, ${\rm tr}_B(\hat\rho_{\rm ME})=\hat 1/d$.

\subsubsection{Flip-type operators}	
	Another example provides a deeper insight into the symmetry of the ME states.
	For this reason, let us consider the so-called flip operator, $\hat F|x,y\rangle=|y,x\rangle$, which exchanges the two subsystems.
	More generally, we study a transformed version,
	\begin{align}
		\hat L=(\hat A\otimes \hat B)\hat F(\hat A\otimes \hat B)^\dagger,
	\end{align}
	for arbitrary operators $\hat A$ and $\hat B$.
	This kind of operator has been intensively studied in Ref.~\cite{MW09} for characterizing RU channels.

	The operator $\hat L$ maps a state $|x,y\rangle$ as follows:
	\begin{align}\label{eq:sepmapflip}
		\hat L|x,y\rangle=\hat A\hat B^\dagger|y\rangle\otimes\hat B\hat A^\dagger|x\rangle.
	\end{align}
	Hence, it is convenient to consider the singular-value decomposition $\hat B\hat A^\dagger=\hat U_1\hat \Sigma\hat U_2^\dagger$, with $\hat \Sigma$ being the diagonal matrix of decreasing singular values, $\Sigma_0\geq\dots\geq\Sigma_{d-1}\geq0$ and two unitary operators $\hat U_{1}$ and $\hat U_2$.
	Inserting this decomposition, Eq.~\eqref{eq:sepmapflip} can be rewritten in the form
	\begin{align}\label{eq:mappingFlip}
		\hat L|x,y\rangle=(\hat U_2\otimes\hat U_1)(\hat\Sigma\otimes\hat\Sigma)\hat F(\hat U_2^\dagger\otimes\hat U_1^\dagger)|x,y\rangle.
	\end{align}
	As $\hat U_{2(1)}$ is a unitary basis transformation of the first (second) mode not affecting eigenvalues, we may simplify the problem by choosing $\hat U_1=\hat U_2=\hat 1$ from now on.
	Now, the spectral decomposition reads
	\begin{align}\label{eq:spectdecFlip}
		(\hat \Sigma \otimes \hat \Sigma)\hat F=&\sum_{m}\Sigma_m^2|m,m\rangle\langle m,m|
		\\\nonumber&{+}\sum_{m<n}\Sigma_m\Sigma_n
		\left(|\psi_{mn}^{+}\rangle\langle\psi_{mn}^{+}|-|\psi_{mn}^{-}\rangle\langle\psi_{mn}^{-}|\right),
	\end{align}
	where the eigenvectors $|\psi_{mn}^{\pm}\rangle=(|m,n\rangle \pm |n,m\rangle)/\sqrt 2$ form symmetric (spanned by $|\psi_{mn}^{+}\rangle$ together with $|m,m\rangle$) and skew-symmetric subspaces (spanned by $|\psi_{mn}^{-}\rangle$).
	We deduce the upper and lower bound for all states,
	\begin{align}
		 g^{\max} =\Sigma_0^2
		 \text{ and }
		 g^{\min} =-\Sigma_0 \Sigma_1.
	\end{align}
	
	For separable states, we have the bounds
	\begin{align}
		g_{\rm S}^{\max}=\Sigma^2_0
		\text{ and }
		g_{\rm S}^{\min}=0,
	\end{align}
	which are given from the partial transposition $(|\Phi\rangle\langle\Phi|)^{T_B}=\hat F$ and the approach in Sec.~V of Ref.~\cite{SV09}.
	Comparing these bounds for separable states with the bounds for all states, we see that the upper bounds coincide, $g_{\rm S}^{\max}=g^{\max}$.
	Hence, as long as there are at least two nonvanishing singular values $\Sigma_0\Sigma_1\neq 0$, only the lower bound $g^{\min}_{\rm S}> g^{\min}$ provides us with a reasonable test for inseparable states. 

	Let us now consider ME states.
	Applying our optimization equations and following the same procedure as above, one gets
	\begin{align*}
		&\hat L[\hat 1\otimes\hat U]|\Phi\rangle
		{=}\hat 1\otimes \hat \Sigma\hat U^T\hat\Sigma|\Phi\rangle
		{=}\hat 1\otimes (d\hat\Gamma\hat U)|\Phi\rangle.
	\end{align*}
	Hence, we have $\hat\Gamma=\hat \Sigma\hat U^T\hat \Sigma\hat U^\dagger/d$ and
	\begin{align}\label{eq:flipPrelim}
		g^{\rm opt}_{\rm ME}=&\frac{1}{d}{\rm tr}(\hat\Sigma\hat U^T\hat\Sigma\hat U^\dagger)=\frac{1}{d}{\rm tr}(\hat U^\ast\hat\Sigma\hat U\hat\Sigma).
	\end{align}
	The maximal expectation value is given for ME states in the symmetric subspace, which is spanned by ${|\psi^{+}_{m,n}\rangle=(|m,n\rangle+|n,m\rangle)/\sqrt{2}}$ for $m\leq n$.
	Equivalently, this means that $\hat U=\hat U^T$, which simplifies Eq.~\eqref{eq:flipPrelim} to
	\begin{align}
		g^{\rm opt}_{\rm ME}=&\frac{1}{d}\sum_{m,n=0}^{d-1} \Sigma_m\Sigma_n |\langle n|\hat U|m\rangle|^2,
		\\\text{i.e., }g^{\max}_\mathrm{ME}=&\frac{1}{d}\sum_{m=0}^{d-1} \Sigma_m^2,
	\end{align}
	where the latter maximum again follows from Chebyshev's sum inequality in Appendix~\ref{app:Cheb}.

	For the minimum $g_{\rm ME}^{\min}$, we proceed similarly considering the cases of even and odd dimensionality $d$ separately.
	From the spectral decomposition given by Eq.~\eqref{eq:spectdecFlip}, one can see that the generalized eigenvector $|\psi_{\rm ME}\rangle=d^{-1/2}\hat 1\otimes\hat U|\Phi\rangle$ should be an element of the ${d(d-1)/2}$-dimensional skew-symmetric subspace spanned by ${|\psi^{-}_{m,n}\rangle=(|m,n\rangle-|n,m\rangle)/\sqrt{2}}$ for $m<n$.
	Equivalently, this means that $\hat U$ should be, in the case of an even $d$, an antisymmetric operator, $\hat U^T=-\hat U$.
	Thus, we find
	\begin{align}\label{eq:minEVMEfliptype}
		g_{\rm ME}^{\min}=-\frac{1}{d}\sum_{m,n=0}^{d-1} \Sigma_m\Sigma_n|\langle n|\hat U|m\rangle|^2.
	\end{align}
	In order to find the tight lower bound, we utilize the Youla (or Slater) decomposition of a skew-symmetric operator of even dimension, $\hat U=\hat V\hat J\hat V^T$, where $\hat V$ is unitary and $\hat J=\sum_{n=0}^{(d-2)/2} [J_n(|2n\rangle\langle 2n+1|-|2n+1\rangle\langle 2n|)]$ is a block-diagonal skew-symmetric matrix~\cite{DY61}.
	In our case, we have $J_n=1$, which is the only choice that allows $\hat U$ to be unitary.
	Setting $\hat V =\hat 1$ yields the desired minimum in Eq.~\eqref{eq:minEVMEfliptype}.

	For the odd case, one can add the minimal positive eigenvalue $\Sigma_{d-1}^2$ of $\hat L$ yielding the smallest possible positive contribution to $g_{\rm ME}^{\min}$ and preserving the unitarity of $\hat U$.
	In detail, we modify our optimal $\hat U=\hat J$ for the even case such that $\hat U=\hat J + |d-1\rangle\langle d-1|$ for odd $d$.
	In conclusion, we obtain
	\begin{align}
		g_{\rm ME}^{\min}=-\frac{1}{d}\left\lbrace\begin{array}{ll}2
			\sum\limits_{n=0}^{(d-2)/2}\Sigma_{2n}\Sigma_{2n+1} & \text{for }d\text{ even,}\\
			2\sum\limits_{n=0}^{(d-3)/2}\Sigma_{2n}\Sigma_{2n+1}-\Sigma_{d-1}^2 & \text{for }d\text{ odd.}\\
		\end{array}\right.
	\end{align}
	Again, we need to examine the eligibility of $\hat L$ to witness ME states by checking $g_{\rm{ME}}^{\max}$ against $g^{\max}$ and $g_{\rm{ME}}^{\min}$ against $g^{\min}$, respectively.
	Given that there are at least two different non-vanishing singular values $\Sigma_0\neq\Sigma_1\neq 0$, the upper bounds do not coincide, $g_{\rm{ME}}^{\max}<g^{\max}$.
	For the lower bounds, we notice that they differ, $g_{\rm{ME}}^{\min}>g^{\min}$, under the premise that $d\geq 3$ and that there are at least two nonvanishing singular values, $\Sigma_0\Sigma_1\neq 0$.
	Thus, both upper and lower bounds can be employed as a test for non-ME states. 

\subsubsection{Observations}
	From these very first examples for constructing ME probes [see Eq.~\eqref{eq:MEwitness}], we see that our optimization approach in terms of the generalized eigenvalue equation~\eqref{eq:ME-EValEq} is a useful technique to construct witnesses for ME states.
	Known results could be easily derived, generalized, and compared to a related approach for separable states.
	Comparing the above solutions, one can even find a remarkable feature that relates to our geometric considerations in the previous section.
	Namely, the values for $g_{\rm ME}^{\max}$ or $g_{\rm S}^{\max}$ are closely related to $1$-norm or $\infty$-norm systems, respectively.
	In the following, we will exploit this observation in more detail.

%-----------------------------------------------------------------------------------------------
%-----------------------------------------------------------------------------------------------
%-----------------------------------------------------------------------------------------------
\section{Complementary Schmidt decomposition}\label{sec:pure}
	In this last section, we will apply our witnessing approach to study Hermitian rank-one operators $\hat L=|\psi\rangle\langle\psi|$.
	Based on our approach and the previously performed studies on entanglement, we are able to assess the entanglement properties of $|\psi\rangle$.
	Finally, we will construct the complementary Schmidt decomposition.

%-----------------------------------------------------------------------------------------------
\subsection{Rank one witnesses}\label{subsec:rankOne}
	As pointed out before (see also Appendix~\ref{app:states}), any state $|\psi\rangle$ can be written as $|\psi\rangle=\hat 1\otimes\hat M|\Phi\rangle$.
	Inserting this into our optimization given by Eq.~\eqref{eq:ME-EValEq} and performing the same algebra as done in the previous examples, we obtain
	\begin{align}\label{eq:PureGamma}
	\begin{aligned}
		\hat \Gamma=\frac{{\rm tr}(\hat M^\dagger\hat U)}{d}\hat M\hat U^\dagger\\
		\text{and }
		g^{\rm opt}_{\rm ME}=\frac{1}{d}\left|{\rm tr}(\hat M^\dagger\hat U)\right|^2.
	\end{aligned}
	\end{align}
	As local unitaries affect neither separability nor the ME property, we directly start from the Schmidt decomposition~\eqref{eq:SchmidtDec} in a rotated computational basis, i.e., $|e_m,f_n\rangle=|m,n\rangle$.
	In particular, this means that $\hat M$ is the diagonal matrix of Schmidt coefficients,
	\begin{align}
		\hat M={\rm diag}(\sigma_0,\ldots,\sigma_{d-1})=\mathrm{diag}(\vec \sigma).
	\end{align}
	Thus, we get ($\hat U=\hat 1$)
	\begin{align}\label{eq:MEpureMax}
		g^{\max}_{\rm ME}=\frac{1}{d}\left|\sum_{n=0}^{d-1}|\sigma_n|\right|^2=\frac{\|\vec\sigma\|_{1}^2}{d}.
	\end{align}
	Again, this can be compared with the separability eigenvalue approach,
	\begin{align}\label{eq:SpureMax}
		g_{\rm S}^{\max}=&\max\{|\sigma_n|^2\}_{n=0,\dots,d-1}=\|\vec \sigma\|_\infty^2,
	\end{align}
	see Sec.~IV~A in Ref.~\cite{SV09}.

\begin{figure}
		\includegraphics[width=4cm]{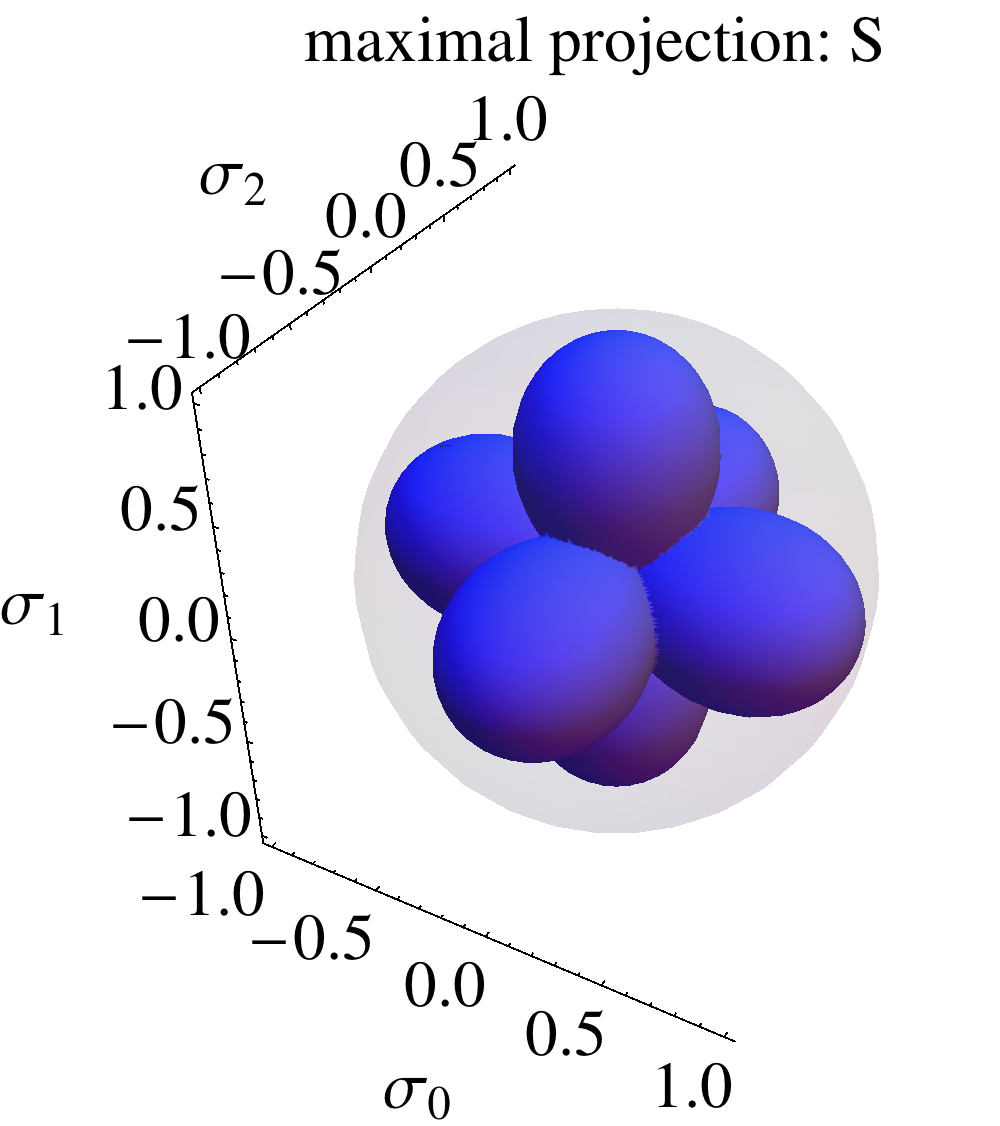}
		\includegraphics[width=4cm]{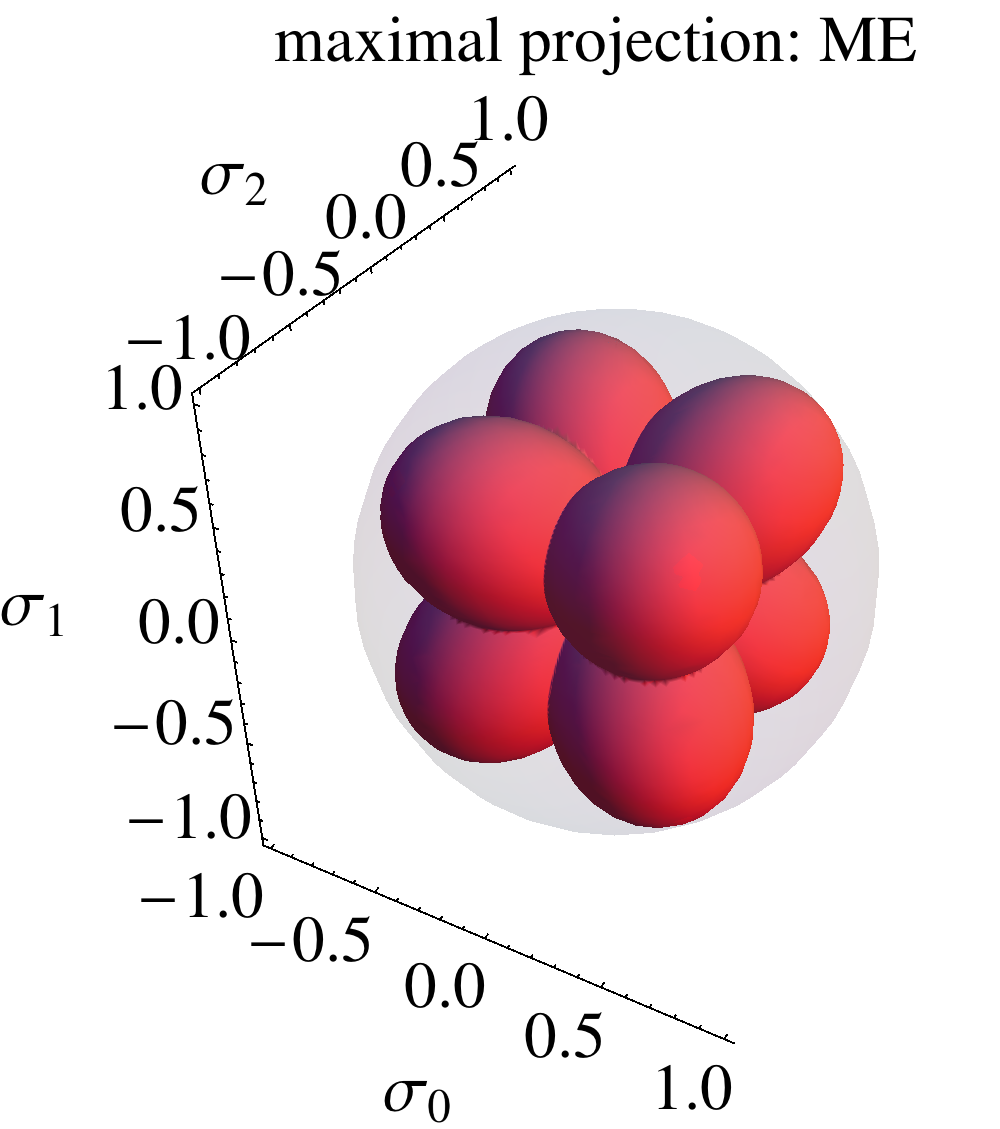}
		\caption{(Color online)
			The maximal projection of a pure bipartite state $|\psi\rangle$ with the Schmidt coefficients $\vec \sigma$ onto separable (left plot) and ME (right plot) is shown.
			The (gray) sphere indicates the normalization of the state, $\|\vec \sigma\|^2_2=1$.
			Whenever the overlap $\langle\psi|\hat\varrho|\psi\rangle$ of a bipartite state is outside of one of those surfaces, we have an inseparable (left) or non-ME (right) state $\hat\varrho$.
		}\label{fig:overlay}
	\end{figure}

	In Fig.~\ref{fig:overlay}, we plot, for $d=3$ and for real-valued singular value vectors $\vec \sigma\in\mathbb R^3$, the bounds in Eqs.~\eqref{eq:MEpureMax} and~\eqref{eq:SpureMax}.
	The left panel shows $\|\vec\sigma\|_\infty^2\vec\sigma$ for normalized states $|\psi\rangle$, $\|\vec \sigma\|^2_2=1$.
	Correspondingly, the right panel depicts $d^{-1}\|\vec\sigma\|_1^2\vec\sigma$.
	This means the bound $g_{\rm S}^{\max}$($g_{\rm ME}^{\max}$) is, in the left(right) panel, the distance of the surface to $(0,0,0)^T$ in the $\vec \sigma$ direction.

	Let us consider an application of such rank-one test operator.
	We find that a quantum state $\hat\varrho$ is neither separable nor ME, if for the fidelity with the state $|\psi\rangle$ the inequality
	\begin{align}\label{eq:exampleRankOne}
		\langle \psi|\hat\varrho|\psi\rangle>\max\{g_{\rm ME}^{\max},g_{\rm S}^{\max}\}
	\end{align}
	holds.
	For instance, we may take the state
	\begin{align}
		\hat\varrho=(1-p)\frac{1}{d^2}\hat 1\otimes \hat 1+p|\psi\rangle\langle \psi|,
	\end{align}
	which is a mixture of a pure state and white noise.
	Note that the normalized identity, i.e. the white-noise contribution, is both separable as well as ME.
	Now we can estimate, by condition~\eqref{eq:exampleRankOne}, the maximal amount of white noise, $1-p$, that the system can undergo without losing its entanglement or its non-ME property.
	This holds for all
	\begin{align}
		p>\frac{d^2\max\{\|\vec\sigma\|_1^2/d,\|\vec \sigma\|_\infty^2\}-1}{d^2-1}.
	\end{align}
	From the geometric point of view, this means that $\langle\psi|\hat\varrho|\psi\rangle$ is outside the surfaces in Fig.~\ref{fig:overlay}.

%-----------------------------------------------------------------------------------------------
\subsection{Discrete Fourier transform and decompositions in terms of ME states}
	In the case of separability, it has been shown in Ref.~\cite{SV09} that the nontrivial solutions, $|a,b\rangle$ for $g_{\rm S}^{\rm opt}\neq 0$, of the optimization equations~\eqref{eq:SEValEq} for $\hat L=|\psi\rangle\langle\psi|$ give the Schmidt decomposition of $|\psi\rangle$.
	Here we will search for a similar possibility.
	Hence, let us reevaluate the solution in Eq.~\eqref{eq:PureGamma}.

	Suppose we have found a set of solutions $\{\hat U_k\}_{k=0,\dots,d-1}$.
	For simplicity, we consider only such unitaries that commute with $\hat M={\rm diag}(\vec \sigma)$.
	Hence, we can write
	\begin{align}\label{eq:commutingAnsatz}
		\hat U_k={\rm diag}(\exp[i\varphi_{k,0}],\dots,\exp[i\varphi_{k,d-1}]),
	\end{align}
	and obtain a diagonal $\hat \Gamma_k$ operator in Eq.~\eqref{eq:PureGamma}.
	Finally, our ansatz for expanding $|\psi\rangle$ is
	\begin{align}
		|\psi\rangle=\sum_{n=0}^{d-1} \sigma_n|n,n\rangle=\sum_{k=0}^{d-1} \tau_k\frac{1}{\sqrt d}\sum_{n=0}^{d-1}e^{i\varphi_{k,n}}|n,n\rangle.
	\end{align}
	Hence we have a system of equations ($n=0,\dots,d-1$):
	\begin{align}\label{eq:DFT1}
		\sigma_n=\sum_{k=0}^{d-1}\frac{1}{\sqrt d}e^{i\varphi_{k,n}}\tau_k.
	\end{align}

	At least one solution can be identified when taking
	\begin{align}\label{eq:DFT2}
		\varphi_{k,n}=-\frac{2\pi}{d}kn-\vartheta_k.
	\end{align}
	The unique character of such a choice is the fact that Eqs.~\eqref{eq:DFT1} for $n=0,\dots,d-1$ with the phases in Eq.~\eqref{eq:DFT2} describe a Fourier transform.
	Namely, we have
	\begin{align}\label{eq:GDFT}
		\tau_k=\frac{e^{i\vartheta_k}}{\sqrt d}\sum_{n=0}^{d-1}e^{2\pi i kn/d}\sigma_n,
	\end{align}
	where we choose $\vartheta_k$ such that $\tau_k\geq0$.
	The transformation in Eq.~\eqref{eq:GDFT} can be called a generalized discrete Fourier transform (GDFT) which maps the non-negative vectors $\vec \sigma=(\sigma_0,\dots,\sigma_{d-1})^T$ to the non-negative vectors $\vec \tau=(\tau_0,\dots,\tau_{d-1})^T$.
	Denoting the states ${|\psi_{\mathrm{ME},k}\rangle=\hat 1\otimes\hat U_k|\Phi\rangle/\sqrt d=|\mathcal F_{k,0}\rangle}$, we can write
	\begin{align}\label{eq:CSDec}
		|\psi\rangle=\sum_{k=0}^{d-1}\tau_k|\mathcal F_{k,0}\rangle,
	\end{align}
	where $\hat U_k$ in Eq.~\eqref{eq:commutingAnsatz} is defined by the phases in Eq.~\eqref{eq:DFT1}.
	Note that our choice is also an orthonormal decomposition $\langle \mathcal F_{k,0}|\mathcal F_{k',0}\rangle=\delta_{k,k'}$ (see also Appendix~\ref{app:Fourier}).

	Hence, one way to represent a state in terms of ME states has been found.
	The remarkable aspect of the form~\eqref{eq:CSDec} is that the coefficients $\vec \tau=(\tau_0,\dots,\tau_{d-1})^T$ of the expansion in terms of ME states are given by the GDFT of the (standard) Schmidt decomposition in terms of separable states.
	Therefore, we may refer to the expansion~\eqref{eq:CSDec} as the {\it complementary Schmidt decomposition}.

	In Appendix~\ref{app:Fourier}, it is shown for the discrete Fourier transform $\boldsymbol F$ that a vector $\vec \sigma$ with non-negative entries has the image $\vec\tau'=\boldsymbol F\vec\sigma$ for which $\|\vec \tau'\|_{\infty}=d^{-1/2}\|\vec\sigma\|_1$ holds.
	Because $\tau_k=e^{i\vartheta_k}\tau'_k$, we get the same result for $\vec \tau$.
	Analogously, we conclude from the inverse GDFT that $\|\vec\sigma\|_\infty=d^{-1/2}\|\vec\tau\|_1$.
	Note that the inverse GDFT may be computed similarly to the ansatz presented in this section starting from the complementary Schmidt decomposition and the maximally non-ME (separable) states.
	In addition, we get identical $2$-norms for $\vec \sigma$ and $\vec \tau$ (see also Appendix~\ref{app:Fourier}).
	In summary, the GDFT yields the following important relations between the Schmidt and complementary Schmidt coefficients:
	\begin{align}
		\|\vec \tau\|_2=\|\vec \sigma\|_2,\text{ }
		\|\vec \tau\|_\infty=\frac{\|\vec \sigma\|_1}{\sqrt d},
		\text{ and }\|\vec \tau\|_1=\sqrt d\|\vec \sigma\|_\infty.
	\end{align}
	This highlights the dual character of the complementary Schmidt decomposition.

	Moreover, a maximally non-ME (i.e., separable) state is described in terms of equally weighted complementary Schmidt coefficients $\vec\tau=d^{-1/2}(1,\dots,1)^T$.
	Up to unitary transformations $\hat V_A$ and $\hat V_B$, we have, for any separable state $|\psi_{\rm S}\rangle=|a,b\rangle$,
	\begin{align}
		|0,0\rangle=\hat V_A\otimes\hat V_B|a,b\rangle=\sum_{k=0}^{d-1}\frac{1}{\sqrt d}|\mathcal F_{k,0}\rangle,
	\end{align}
	keeping in mind that $\{|\mathcal F_{k,0}\rangle\}_{k=0,\dots,d-1}$ are orthonormal ME states.
	For the same reasons, any ME state takes the form
	\begin{align}
		|\psi_\mathrm{ME}\rangle=\hat V_A^\dagger\otimes\hat V_B^\dagger\sum_{k=0}^{d-1}\delta_{k,0}|\mathcal F_{k,0}\rangle.
	\end{align}

\begin{figure}
		\includegraphics[width=8.5cm]{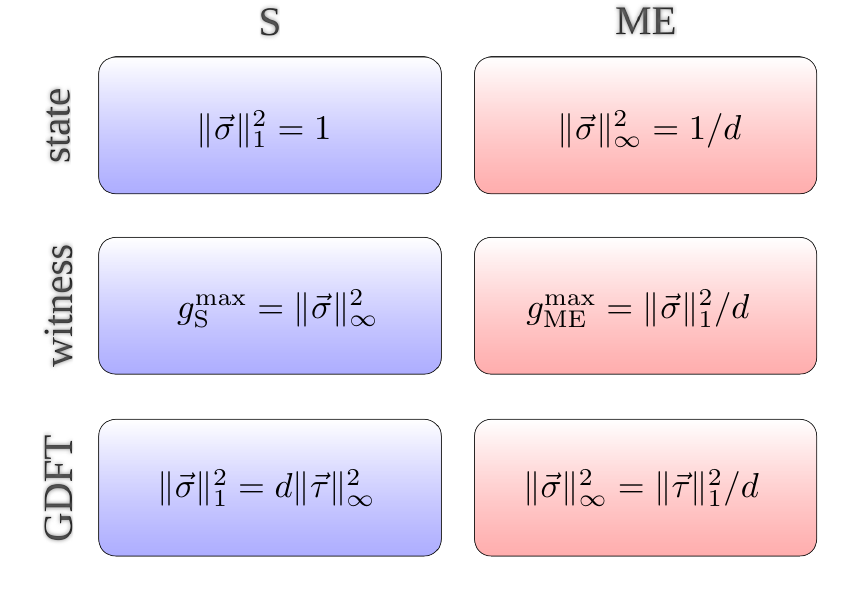}
		\caption{(Color online)
			The interpretations of norms of the vector of Schmidt coefficients $\vec \sigma$ are shown.
			For normalized states, we have $\|\vec\sigma\|_2=1$.
			If, in addition, the top row is fulfilled, we have a separable (S) or ME state.
			Choosing a pure state to be a witness, we get the upper bounds for separable or ME states in the middle row.
			The relation between the Schmidt coefficients and complementary Schmidt coefficients $\vec\tau$, with $\|\vec\tau\|_2=1$, under the GDFT [Eq.~\eqref{eq:GDFT}] is given in the bottom row.
		}\label{fig:Relation}
	\end{figure}

	In Fig.~\ref{fig:Relation}, we summarize the complementary relations between ME and separable states.
	Any pure normalized state $\hat\varrho=|\psi\rangle\langle\psi|$ ($\|\vec\sigma\|_2=\|\vec\tau\|_2=1$) is characterized by the vector of Schmidt coefficients $\vec\sigma$ or its GDFT-mapped coefficients $\vec\tau$.
	One result from Sec.~\ref{subsec:geoState} is given in the first row and can be extended with the results in the third row of Fig.~\ref{fig:Relation}.
	That is, such a state $\hat\varrho$ is a separable or ME state if and only if $\|\vec \sigma\|_1=d^{1/2}\|\vec\tau\|_\infty=1$ or $d^{1/2}\|\vec \sigma\|_\infty=\|\vec\tau\|_1=1$, respectively.
	Test operators of the form $\hat L=|\psi\rangle\langle\psi|$ are in the dual space of density operators.
	Hence, the bounds for ME and separable states are expressed in the complementary form (see Sec.~\ref{subsec:rankOne}).
 	From the rows 2 and 3 of Fig.~\ref{fig:Relation}, we consequently get $g_{\rm ME}^{\max}=\|\vec \sigma\|_1^2/d=\|\vec \tau\|_\infty^2$ or $g_{\rm S}^{\max}=\|\vec \tau\|_1^2/d=\|\vec \sigma\|_\infty^2$.
	Finally, the maximal expectation value of $\hat L$ for arbitrary quantum states is given by the only nonzero eigenvalue, $g^{\max}=\|\vec\sigma\|_2^2=\|\vec\tau\|_2^2=1$.

%-----------------------------------------------------------------------------------------------
%-----------------------------------------------------------------------------------------------
%-----------------------------------------------------------------------------------------------
\section{Conclusions}\label{sec:summary}
	In summary, we exploited the relations between quantum channels and bipartite quantum states.
	We derived a method that allowed us to construct the corresponding witnesses.
	We performed a full analytical characterization of pure states, e.g., for estimating the amount of imperfections a quantum property can withstand.

	In a first step, we identified two maximal quantum channels, random unitary and random projective channels.
	Random unitaries are completely characterized by a deterministic evolution and classical statistics.
	The complement was introduced as a random projective channel.
	These kinds of processes are governed by the quantum measurement-induced state collapse.
	Its relation to entanglement-breaking channels was discussed.
	Applying the Choi-Jamio\l{}kowski isomorphism, we could show that these channels are mapped onto two complementary forms of bipartite quantum states.
	The isomorphism acts in an anticorrelated way:
	The quantum-dominated projective channels are transformed into classically correlated states,
	and deterministic unitary channels are mapped onto maximally entangled states.

	In a second step, a technique for constructing witnesses was derived to probe nonrandom unitary channels or, equivalently, nonmaximally entangled states.
	The resulting, generalized eigenvalue equations have been compared with a related approach to uncover inseparable states or, equivalently, nonprojective quantum channels.
	Some examples underlined the general functionality of our method.
	With a single observable, one can perform a joint witnessing of inseparable and nonmaximally entangles states.
	Moreover, the computed bounds for the witnessing are tight.

	An example of particular interest was formulated in terms of rank-one operators being defined by a single pure state.
	We showed that the maximal overlap of this state with maximally entangled ones is given in terms of the $1$-norm of the vector of Schmidt coefficients, whereas the maximal fidelity with product states is given by its $\infty$-norm.
	Finally, we introduced a complementary Schmidt decomposition.
	Contrary to the standard expansion with orthonormal separable states, the complementary Schmidt decomposition expands a state in terms of maximally entangled ones.
	In particular, the Schmidt coefficients and the complementary Schmidt coefficients are connected via a discrete Fourier transform.

	In conclusion, our method is useful to characterize channels and states of a classical or quantum character in a unified manner.
	Because witnesses define tangent hyperplanes to the studied convex sets, the presented approach allows one to identify the full geometry.
	Some steps in this direction have been done in the present work.
	Moreover, our criteria, e.g., in terms of the presented correlation measurements, are directly applicable in present experiments.

%-----------------------------------------------------------------------------------------------
%-----------------------------------------------------------------------------------------------
%-----------------------------------------------------------------------------------------------
\section*{Acknowledgements}
	The authors acknowledge support by the Deutsche Forschungsgemeinschaft through SFB 652/3.

%-----------------------------------------------------------------------------------------------
%-----------------------------------------------------------------------------------------------
%-----------------------------------------------------------------------------------------------
\appendix
%-----------------------------------------------------------------------------------------------
\section{Relating operators and states}\label{app:states}
	In this appendix, we will provide some simple, yet useful relations for the representation of bipartite pure states.
	First, any pure state can be written as
	\begin{align}\label{eq:PureStateRep}
		|\psi\rangle=\hat 1\otimes\hat M|\Phi\rangle=\hat M^T\otimes\hat 1|\Phi\rangle,
	\end{align}
	with $|\Phi\rangle=\sum_{q}|q,q\rangle$ and $\hat M=\sum_{m,n}\psi_{m,n}|n\rangle\langle m|$.
	Inserting $\hat M$ yields the proper decomposition of the state $|\psi\rangle=\sum_{m,n}\psi_{m,n}|m,n\rangle$.
	Second, we have
	\begin{align}
		\hat A\otimes\hat B|\Phi\rangle=\hat A\hat B^T\otimes\hat 1|\Phi\rangle=\hat 1\otimes \hat B\hat A^T|\Phi\rangle.
	\end{align}
	Expanding both operators in the computational basis, $\hat A=\sum_{m,n} A_{m,n}|m\rangle\langle n|$ and $\hat B=\sum_{m',n'} B_{m',n'}|m'\rangle\langle n'|$, proves this relation.
	Third, the following equivalence obviously holds:
	\begin{align}
		\hat M=\hat N \Leftrightarrow
		\hat 1\otimes\hat M|\Phi\rangle=\hat 1\otimes\hat N|\Phi\rangle,
	\end{align}
	which is a direct consequence of the identity~\eqref{eq:PureStateRep}.
	Finally, one can directly evaluate that
	\begin{align}
		\langle\Phi|\hat 1\otimes\hat M|\Phi\rangle={\rm tr}(\hat M).
	\end{align}

%-----------------------------------------------------------------------------------------------
\section{Chebyshev's sum inequality}\label{app:Cheb}
	In this appendix, we generalize the proof of Chebyshev's sum inequality for our needs.
	The standard form of this inequality states that for two ordered sequences, $a_0\geq\dots\geq a_{d-1}$ and $b_0\geq\dots\geq b_{d-1}$, one has
	\begin{align}\label{eq:ChebSum}
		\frac{1}{d}\sum_{n=0}^{d-1}a_nb_n\geq\left(\frac{1}{d}\sum_{n=0}^{d-1} a_n\right)\left(\frac{1}{d}\sum_{n=0}^{d-1} b_n\right),
	\end{align}
	and if one (and only one) of the sequences has the inverse ordering, then the inequality~\eqref{eq:ChebSum} changes the relation to ``$\leq$''.

	In addition to this well-known form, let us assume a symmetric matrix of non-negative elements,
	\begin{align}
		G_{i,j}=G_{j,i}=|\langle i|\hat U|j\rangle|^2 
	\end{align}
	for a unitary operator $\hat U$.
	Suppose we have two finite and sorted sequences of numbers, $a_0\leq\ldots\leq a_{d-1}$ and $b_0\leq\ldots\leq b_{d-1}$.
	It then holds that
	\begin{align*}
		0\leq S=\sum_{i,j=0}^{d-1} G_{i,j}(a_i-a_j)(b_i-b_j),
	\end{align*}
	with $(a_i-a_j)(b_i-b_j)\geq0$ for all $i,j$.
	The expansion of the latter expression yields
	\begin{align*}
		S{=}\!\sum_{i,j} G_{i,j} a_ib_i{+}\!\sum_{i,j} G_{i,j} a_jb_j
		{-}\!\sum_{i,j}G_{i,j} a_ib_j{-}\!\sum_{i,j}G_{i,j} a_jb_i.
	\end{align*}
	Replacing $i\leftrightarrow j$ in the second and fourth terms and using the symmetry of $G_{i,j}$, we get
	\begin{align*}
		S=2\sum_{i}\Big[\sum_jG_{i,j}\Big] a_ib_i-2\sum_{i,j}G_{i,j}a_ib_j\geq0.
	\end{align*}
	This can be rewritten in the form of a generalized Chebyshev's sum inequality,
	\begin{align}
		\sum_{i,j=0}^{d-1}G_{i,j}a_ib_j\leq \sum_{i=0}^{d-1} a_ib_i,
	\end{align}
	where we took into account that for any unitary $\hat U$, one has $\sum_{j} G_{i,j}=\sum_j \langle i|\hat U|j\rangle\langle j|\hat U^\dagger|i\rangle=\langle i|\hat U\hat 1\hat U^\dagger|i\rangle=1$.

	In addition, we may define the unitary permutation operation $\hat U_{0}|j\rangle=|d-1-j\rangle$ and $G'_{i,j}=|\langle i|\hat U\hat U_0|j\rangle|^2=G_{i,d-1-j}$ for $\hat U'=\hat U\hat U_0$.
	Now we have
	\begin{align*}
		0\geq S'=\sum_{i,j=0}^{d-1} G_{i,j}(a_i-a_j)(b_{d-1-i}-b_{d-1-j}).
	\end{align*}
	Similarly to the proof above, one can show that
	\begin{align}
		\sum_{i,j=0}^{d-1}G_{i,j}a_ib_{d-1-j}=\sum_{i,j=0}^{d-1}G'_{i,j}a_ib_{j}\geq \sum_{i=0}^{d-1} a_ib_{d-1-i}
	\end{align}
	holds for any unitary $\hat U$.

%-----------------------------------------------------------------------------------------------
\section{Maximally entangled Fourier basis}\label{app:Fourier}
	The standard computational basis $\{|m,n\rangle\}_{m,n=0,\dots,d-1}$ is a product basis.
	In this section, an orthonormal basis of ME states is given.
	It reads as
	\begin{align}
		&|\mathcal F_{m,n}\rangle=\frac{1}{\sqrt d}\sum_{q=0}^{d-1}\omega^{qm}|q,q+n \text{ mod }d\rangle,
		\\&\text{for }\omega=\exp(2\pi i/d)
	\end{align}
	and $m,n=0,\dots,d-1$.
	The expansion coefficients represent the kernel of the discrete Fourier transform and the state of the second mode relates to the discrete convolution.

	We may also define the two unitary operators $\hat U=\sum_{q=0}^{d-1}\omega^q|q\rangle\langle q|$ and $\hat V=\sum_{q=0}^{d-1} |q+1\text{ mod }d\rangle\langle q|$ to see that these states are ME states:
	\begin{align}
		|\mathcal F_{m,n}\rangle=\hat U^m\otimes\hat V^n \frac{1}{\sqrt d}\sum_{q=0}^{d-1} |q,q\rangle.
	\end{align}
	With this form, we can also verify that these states are $d^2$ orthonormal ones and, thus, form an orthonormal basis.
	We have
	\begin{align*}
		\langle \mathcal F_{m,n}|\mathcal F_{m',n'}\rangle=\frac{1}{d}\langle \Phi|\hat U^{(m'-m)}\otimes\hat V^{(n'-n)}|\Phi\rangle.
	\end{align*}
	In the case $n\neq n'$, the resulting sum is empty.
	The missing part is given by the geometric series relation: $\sum_{q=0}^{d-1} (\omega^{m-m'})^q=d\delta_{m-m',0}$.
	We additionally have
	\begin{align}
		\hat 1\otimes\hat 1=\sum_{m,n=0}^{d-1}|\mathcal F_{m,n}\rangle\langle \mathcal F_{m,n}|.
	\end{align}
	Finally, let us retrieve the computational separable basis in a form of maximally non-ME states,
	\begin{align}
		|p,q\rangle=\frac{1}{\sqrt d}\sum_{m=0}^{d-1}\omega^{-mp}|\mathcal F_{m,n-p+q}\rangle.
	\end{align}

	In addition, let us provide a relation between the discrete Fourier transform and norms.
	Let $\vec x=(x_0,\dots,x_{d-1})^T$ be a $d$-dimensional vector of non-negative numbers, written as $\vec x\geq 0$, and the transformed one is $\vec y=(y_0,\dots,y_{d-1})^T$, i.e.,
	\begin{align}
		\vec y=\boldsymbol F\vec x,
		\text{ with }\boldsymbol F=\frac{1}{\sqrt d}(\omega^{mn})_{m,n=0,\dots,d-1}.
	\end{align}
	Parseval's identity of the unitary transformation $\boldsymbol F$ states that
	\begin{align}
		\|\vec x\|_2=\|\vec y\|_2. 
	\end{align}
	Furthermore, the triangle inequality yields
	\begin{align*}
		|y_m|\leq\frac{1}{\sqrt d}\sum_{n=0}^{d-1} |\omega^{mn}|x_n.
	\end{align*}
	Identifying the right-hand side with $y_0=d^{-1/2}\sum_n x_n$, we find
	\begin{align}
		\|\vec y\|_{\infty}=\frac{\|\vec x\|_1}{\sqrt d}.
	\end{align}
	Similarly, we get, for the inverse Fourier transform $\boldsymbol F^\dagger$ and $\vec y\geq0$, the relation
	\begin{align}
		\|\vec x\|_{\infty}=\frac{\|\vec y\|_1}{\sqrt d}.
	\end{align}

%-----------------------------------------------------------------------------------------------
%-----------------------------------------------------------------------------------------------
%-----------------------------------------------------------------------------------------------

\end{document}